\documentclass[amstex,epsf,amssymb,usenatbib]{mn2e}
\usepackage{epsf,graphics}
\usepackage{etex} 
\reserveinserts{30} 
\usepackage{amsmath,amssymb,natbib}
\usepackage{rotating,times,pictex,graphicx,latexsym,color,longtable}
\def\aj{AJ}                   % Astronomical Journal
\def\apj{ApJ}                 % Astrophysical Journal
\def\aap{A\&A}                % Astronomy and Astrophysics
\def\mnras{MNRAS}             % Monthly Notices of the RAS
\def\prl{Phys.~Rev.~Lett.}    % Physical Review Letters
\title[The Structure of Molecular Clouds: III]{The Structure of Molecular
Clouds: III - A link between cloud structure and star formation mode}

\author[J. H. Rowles, D. Froebrich]{Jonathan Rowles $^{1}$\thanks{E-mail:
jr262@kent.ac.uk, } {Dirk Froebrich $^{1}$\thanks{E-mail: df@star.kent.ac.uk}}\\
$^1$Centre for Astrophysics \& Planetary Science, The University of Kent,
Canterbury, Kent CT2 7NH, U.K.} 

\date{Accepted .....
      Received ..... ;
      in original form .....}

\pagerange{\pageref{firstpage}--xxx}
\pubyear{2010}

\begin{document}
\maketitle
\label{firstpage}

\begin{abstract} 

We analyse extinction maps of nearby Giant Molecular Clouds to forge a link
between driving processes of turbulence and modes of star formation. Our
investigation focuses on cloud structure in the column density range above the
self shielding threshold of 1\,mag A$_V$ and below the star formation threshold
-- the regime in which turbulence is expected to dominate.

We identify clouds with shallow mass distributions as cluster forming. Clouds
that form stars in a less clustered or isolated mode show a steeper mass
distribution. Structure functions prove inadequate to distinguish between clouds
of different star formation mode. They may, however, suggest that the turbulence
in the average cloud is governed by solenoidal forcing. The same is found using
the $\Delta$-variance analysis which also indicates that clouds with a clustered
mode of star formation show an enhanced component of compressive driving in the
turbulent field. Thus, while star formation occurs in each cloud, independent of
the turbulent driving mechanism, compressive forcing appears to be associated
with the formation of stellar clusters.

\end{abstract}

\section{Introduction}
\label{intro}

Stars form due to the local collapse of material in molecular clouds. The 
conditions prior to the collapse result from a complex interplay between 
self-gravity, turbulence, magnetic fields and thermodynamics. Understanding the
effects of each of these influences leads to a better knowledge of how stars
form. In this paper we are interested in the role played by interstellar 
turbulence and it's effect on the structure of Giant Molecular Clouds (GMCs). 

The subject of astrophysical turbulence is complex and not fully understood, 
though there are many useful reviews on the topic (e.g. Elmegreen \& Scalo
\citep{2004ARA&A..42..211E}, Scalo \& Elmegreen \citep{2004ARA&A..42..275S},
Brandenburg \&  Nordlund \citep{2009arXiv0912.1340B}). Here we are interested in
comparing the structure of nearby GMCs to investigate whether it has an
influence on the observed star formation properties. Ultimately this might also
be used to determine the nature of the turbulent field, i.e. driven by a
compressive forcing or solenoidal driving (e.g. Federrath et al.
\cite{2010A&A...512A..81F}). 

We will probe the column density structure in nearby GMCs by means of extinction
maps derived from near-infrared observations. This is the least biased way to
estimate column density (Goodman et al. \citep{2009ApJ...692...91G}). In the
first paper in this series (Rowles \& Froebrich \cite{2009MNRAS.395.1640R},
hereafter Paper\,I), we presented new all-sky extinction maps derived using data
from the 2\,Micron All-Sky Survey (2MASS, Skrutskie et al.
\cite{2006AJ....131.1163S}). We used the {\it median} near-infrared colour 
excess technique (NICE) to calculate the extinction (see Lada et al.
\cite{1994ApJ...429..694L}). The nearest 25, 49 and 100 stars to the centre of
each pixel were used, hence the noise can be considered constant throughout the
map. These extinction maps are therefore referred to as {\it con-noise} maps.

In our second paper (Froebrich \& Rowles \cite{2010MNRAS.406.1350F}, hereafter
Paper\,II), we analysed the column density and mass distributions of a selection
of 16 nearby GMCs. To facilitate this we determined new extinction maps using
only the stars within each pixel (i.e. no oversampling), which therefore have a
constant spatial resolution and are referred to as {\it con-res} maps. As a
result of the analysis we found a universal star formation threshold of about
6.0\,$\pm$\,1.5\,mag $A_V$. This threshold separates two different regions in
the clouds. Below the threshold, at low column densities, turbulence dominates
the structure, while at higher column densities gravity is the dominant force.
The low $A_V$ part of the clouds could be fitted by a log-normal distribution.
There were significant differences in the slopes of the column density and mass
distributions when considering only the low $A_V$ regions. This shows that the
properties of the turbulence differ depending on the environment of the cloud.
Regarding the high $A_V$ regions, we found no such differences, implying that
gravity solely dominates these parts. 

Using our extinction maps we can derive column density structure functions
similar to velocity structure functions (e.g. Padoan et al.
\cite{2002ApJ...580L..57P}) for each molecular cloud. This will allow us to
perform a comparison with models of interstellar turbulence. Predictions of
structure functions resulting from a turbulent medium have been presented e.g.
by Schmidt et al. \cite{2008PhRvL.101s4505S}, Kolmogorov 
\citep{1941DoSSR..30..301K}, She \& Leveque \citep{1994PhRvL..72..336S} and 
Boldyrev \citep{2002ApJ...569..841B} (hereafter S08, K41, SL94 and B02, 
respectively). 

In this paper we test for correlations between structure function parameters and
the  properties of the clouds. We also examine the cloud structures using the 
$\Delta$-variance technique (see Stutzki et al. \citep{1998A&A...336..697S} and
Ossenkopf et al. \citep{2008A&A...485..719O}) and in particular the mass
spectral index scaling coefficient. In Sec.\,\ref{sfmeth} we  describe the
methods used for our analysis. In Sec.\,\ref{results} we give the results for
the clouds selected. We discuss these results and give conclusions in
Secs.\,\ref{discuss} and \ref{conclusions}, respectively. 

\section{Method}
\label{sfmeth}

\subsection{Structure functions}
\label{strmet}

The general definition of the structure function is given in 
Eq.\,\ref{struct1}, (e.g. Lombardi et al. \cite{2008A&A...489..143L};  Padoan et
al. \cite{2002ApJ...580L..57P} and  Padoan et al. \cite{2003ApJ...583..308P}).
Here the equation is expressed in terms of our observable -- the column density
or optical extinction $A_V$. 

\begin{equation}
\label{struct1}
S_p (\Delta r) = \left< | A_V \left(r' \right) - A_V \left( r' - \Delta r 
\right) |^p \right>
\end{equation}

$\Delta r$ is the distance between points, $r'$ represents a position in the
map, $A_V$ is the optical extinction at $r'$ (or $r' - \Delta r$) and $p$ is the
order of the structure function. For $p=2$ the equation is the two-point
correlation function of the extinction map. The brackets $\left< \right>$ denote
that the average over all pixel positions $r'$ and all possible directions for
the separation $\Delta r$ of points is applied. For each order $p$ we find the
scaling exponent $s(p)$ by fitting a power-law to values of $S_p (\Delta r)$
against $\Delta r$. This assumes that the scaling exponents are related to the
structure functions by Eq.\,\ref{struct2} (e.g. Padoan et al.
\cite{2003ApJ...583..308P}):

\begin{equation}
\label{struct2}
S_p (\Delta r) \propto \Delta r^{s(p)}.
\end{equation}

Equation\,\ref{struct2} allows us to determine the scaling exponents $s(p)$
which are then normalised to the third order $s(3)$, as a universal behaviour
should be exhibited at low Reynolds numbers (determined by Benzi et al.
\citep{1993PhRvE..48...29B}). 

For each GMC investigated we used the range of spatial scales $\Delta r$ from
0.1\,pc to 1.0\,pc to fit the power law exponent, in order to be able to compare
the results for all clouds. We then follow Padoan et al.
\cite{2002ApJ...580L..57P} and assume the column density scaling exponents
$s(p)/s(3)$ are equivalent to the velocity scaling exponents $\zeta(p)/\zeta(3)$
(Dubrulle \citep{1994PhRvL..73..959D}). They can then be expressed by a relation
of the form as shown in Eq.\,\ref{dubeq}.

\begin{equation}
\label{dubeq}
\frac{\zeta(p)}{\zeta(3)} = (1 - \Delta) \frac{p}{3} + \frac{\Delta}{1 - \beta} 
(1 - \beta^{p/3})
\end{equation}

Here $\beta$ is the intermittency and $\Delta$ is related to the co-dimension
$C$ and intermittency by: $\Delta = C \cdot (1 - \beta)$. The fractal dimension
$D$ of the cloud is related to the co-dimension by $D = 3 - C$. Using this
equation, the K41 relation can be expressed by setting $\Delta $\,=\,0. As
mentioned previously we use the projected column density scaling parameter $s$
instead of the velocity scaling parameter $\zeta$ to keep the same description
of the structure function as in Padoan et al. \cite{2002ApJ...580L..57P}, though
it is not known {\it a priori} whether the parameters correspond to each other.
However, Padoan et al. \cite{2002ApJ...580L..57P} found that the scaling
exponents derived from integrated intensity images follow the velocity scaling
using the B02 relation. 

We may then derive the parameters $\Delta$, $C$, $\beta$ and $D$ for each  cloud
and compare with the velocity field structure functions obtained from  the
aforementioned theoretical works by K41, SL94, B02 and S08. When fitting our
data to Eq.\,\ref{dubeq}, we varied $\Delta$ in the range 0.02 to 1.20 with
increments of 0.01, and $C$ ranged from 0.02 to 3.00 with increments of 0.01 to
find the best match. Only fits with an $rms$ better than 0.1 (in units of
$s(p)/s(3)$) were considered a good fit. This is justified given that
$rms$ is not an absolute measure of the goodness of fit. 

Regarding the previously published models, the simplest to consider is that of 
K41. The energy over a wide range of lengths (known as the `inertial range') in
turbulent flows is redistributed from larger scales into ever smaller scales
until the effects of viscosity become important. In K41, Kolmogorov considered
the structure velocity functions S$_2$ and S$_3$, showing that both are power
laws of the form of Eq.\,\ref{struct2}, with the exponents being 2/3 and 1,
respectively. Due to the self-similarity of turbulence at different scales,
this law can be extended to all powers of $p$ such that $s(p) = p/3$. 

However, experimental measurements for the scaling exponents show a deviation 
from the K41 relation for turbulence, for higher orders of $p$ - known as  {\it
intermittency}. This is exhibited in the non-Gaussian tails when plotting  the
probability density functions (PDFs) of e.g. the column density, and anomalous
scaling of higher order structure functions (e.g. Anselmet et  al.
\citep{1984JFM...140...63A}). Therefore, we need a relation for the scaling that
considers the effects of intermittency. 

SL94 derived a scaling relation that accounts for intermittency. Their relation
shows good agreement when compared to simulations of structures that are not
influenced by magnetic fields. For incompressible turbulence SL94 derived that
$C = 2$ and  $\Delta = 2/3$. Therefore, the SL94 model assumes that $D = 1$,
i.e.  that the most intermittent structures are filaments. 

B02 extended the SL94 relation to model highly supersonic turbulence and also
take account of magnetic fields. They found that $C = 1$ with $\Delta = 2/3$ as
with the SL94 model. Under the B02 model it follows that $D = 2$, i.e. that the
structures are sheet-like in form. 

In S08, the scaling exponents were found to be well described by log-Poisson 
models. In these models $\Delta \simeq 1$, rather than $2/3$. The  modelling
applied two types of forcing to drive the turbulence, solenoidal 
(divergence-free) and compressive (rotation-free). Values for the parameters 
$\Delta$ and $C$ differ depending on the type of forcing applied. Perhaps the 
nature of forcing in real clouds can be found by comparing the observational
data with the S08 model. 

\subsection{The effect of noise on the structure functions}
\label{noistr}

To ascertain the effect of noise in the data when calculating structure
functions from an image we undertook a number of tests. In particular we desired
to know whether (increased) random noise leads to a significant and/or
systematic deviation of $s(p)/s(3)$ compared to a test image without noise, or a
low noise $A_V$ map. In the first instance we made an artificial $A_V$ map of 
400\,$\times$\,400 pixels containing a circle of radius 124 pixels, where the 
pixel values in the circle are varied linearly from 0.81\,mag (edge) to
10.0\,mag (centre). This may be considered analogous to a perfectly spherical
cloud with a smooth density variation from the core to the edge (admittedly, an
unlikely scenario in practice). All other pixels in the map were set to zero.
The structure function of this map happened to correspond to Kolmogorov type
turbulence. We then added 0.28\,mag Gaussian noise (corresponding to the
1\,$\sigma$ noise in our 49$^{th}$ nearest neighbour map -- see Paper\,I) to
this map and fit the parameters $\Delta$, $C$, $\beta$ and $D$. This process was
continued by adding further amounts of noise to the basic map with 0.28\,mag
(1\,$\sigma$) noise. The quantities added were a further 0.0625$\sigma$, 
0.125$\sigma$, 0.1875$\sigma$ and 0.25$\sigma$, generating five images in total.
The results of this exercise are discussed in Sec.\,\ref{noires}. 

In addition to investigating the effect of noise on artificial clouds, we added
random noise to a selection of real clouds. The clouds chosen were Chameleon,
Circinus, Corona Australis, Orion\,A and Orion\,B. We used the nearest 49$^{th}$
nearest neighbour $A_V$ maps and added 0.0625$\sigma$, 0.125$\sigma$, 
0.1875$\sigma$ and 0.25$\sigma$ noise. We performed the structure function 
analysis as with the artificial clouds. These results are also discussed in 
Sec.\,\ref{noires}.

\subsection{The $\Delta$-variance}
\label{delmet}

The $\Delta$-variance method for analysing molecular cloud structures in 
astronomical images (i.e. in two dimensions) was introduced by Stutzki et al.
\cite{1998A&A...336..697S} and improved by Ossenkopf et al.
\cite{2008A&A...485..917O}. The method works by measuring the quantity of 
structure on a particular length scale (e.g. $\Delta r$) and filtering the data
$f(\Delta r)$ with a spherically symmetric `down-up-down' type function. The
function (denoted $\bigodot_l(\Delta r)$), which treats 
different regimes of $\Delta r$ separately, is given by Eq.\,\ref{frehat1} 
below. 

\begin{equation}
 \label{frehat1}
{\bigodot}_l(\Delta r) = {\bigodot}_{l,core}(\Delta r) - {\bigodot}_{l,ann}
(\Delta r)
\end{equation}

The RHS of Eq.\,\ref{frehat1} is defined using Eqs.\,\ref{mexhat1} 
and \ref{mexhat2}.

\begin{equation}
 \label{mexhat1}
{\bigodot}_{l,core}(\Delta r) = \frac{4}{\pi l^2} \exp \left( \frac{\Delta r^2}
{(l/2)^2} \right)
\end{equation}

\begin{equation}
 \label{mexhat2}
\bigodot_{l,ann}(\Delta r) = \frac{4}{\pi l^2 (\nu^2 - 1)} \left[ \exp \left( 
\frac{\Delta r^2}{(\nu l/2)^2} \right) - \exp \left( \frac{\Delta r^2}
{(l/2)^2} \right) \right]
\end{equation}

The $\Delta$-variance, $\sigma^2_\Delta(\Delta r)$ is then defined in 
Eq.\,\ref{delva1}.

\begin{equation}
\label{delva1}
\sigma_{{\Delta}^2}(l) = \left< \left( f(\Delta r) * {\bigodot}_l(\Delta r) 
\right)^2 \right>
\end{equation}

The filter function $\bigodot_l(\Delta r)$ denotes the Fourier transform of  the
filter function with size $l$ and diameter ratio $\nu$. Two particular  filter
types were found to best probe the structure in molecular clouds, a  `Mexican
hat' filter and a `French hat' filter (Ossenkopf et al.
\cite{2008A&A...485..719O}). Both filters treat the core and the annulus 
separately (Eq.\,\ref{frehat1}). The optimal filter function for molecular 
clouds was found to be either a Mexican hat filter with a diameter ratio  of
1.5, (or a French hat filter with a diameter ratio of 2.3 - Ossenkopf et  al.
\cite{2008A&A...485..719O}). We used a Mexican hat filter with a diameter ratio
of 1.5 for our analysis. 

We applied this method to our {\it con-noise maps}. To further investigate the 
dimensions of structures in the clouds (i.e. filaments etc.) we also applied 
this method to the corresponding star density maps. The structure shown in the 
star density maps is enhanced since the extinction in the $A_V$ maps is roughly
proportional to the $\log$ of the star density. Therefore, both the $A_V$ and 
star density maps were used for this analysis. The results are shown in 
Sec.\,\ref{delvrs}. 

Using the $\Delta$-variance technique allows us to also calculate the mass 
spectral index scaling exponent (denoted $\alpha$) of each cloud. This is found
by fitting a power-law to $\sigma^2_\Delta$ against scale. The power law is
fitted over the range of size of the cloud, i.e. between the smallest and
largest scales of the cloud. The smallest scale is determined by eliminating the
pixels affected by oversampling. For the largest scale we chose the point where
the increase in $\sigma^2_\Delta$ begins to tail off before the peak. These
results are also shown in Sec.\,\ref{delvrs}. 

\subsection{The effect of noise on the $\Delta$-variance}
\label{noidvr}

As with the structure functions it is also necessary to consider the effects of
noise on the $\Delta$-variance results. This was investigated in a similar way
as for the structure functions (see Sec.\,\ref{noistr}). The same set of test
images of 400\,$\times$\,400 pixels containing a circle, as well as real clouds
were used. The $\Delta$-variance of these image was calculated. Increments of
Gaussian noise were added to each of the images in exactly the same way as for
the structure function analysis. The results are presented in
Sect.\,\ref{noiresdvar}.

\subsection{Molecular cloud density}
\label{denmet}

We estimate the average physical density of the cloud material in order to
relate it to the cloud properties obtained from our analysis of the column
density and the star formation properties. The average density of material can
be estimated from the mass of the cloud $M$ above a certain $A_V$ threshold, the
number of pixels $N$ the cloud is covering in our $A_V$ map with values above
this threshold, and the distance $d$ to the cloud in the following way:

\begin{equation}
\label{deneq}
\rho [M_{\odot}/pc^3] = \frac{8.78 \cdot 10^{15} \cdot M [M_{\odot}]}
{N^{\frac{3}{2}} \cdot (x [''] d [pc] )^3},
\end{equation}

Here the average density is expressed in solar masses per cubic parsec, the mass
in solar masses, the pixel size $x$ in arcsec and the distance $d$ in parsec.
The power of $3/2$ at the number of pixels of the cloud assumes that the two
visible dimensions in the plane of the sky are a good representation of the
dimension along the line of sight. Note that if one wants to express the density
in units of particles per cubic centimeter, then the constant
8.78\,$\cdot$\,10$^{15}$ needs to be multiplied by a factor of 40.

We determine two average densities for the purpose of this paper. i) The average
density of the entire cloud. This includes all material which is above the self
shielding column density threshold of 1\,mag $A_V$. ii) The average density of
the material above the cloud's star formation threshold. This corresponds to the
density of the material which is potentially involved in star formation. The
results are shown in Sect.\,\ref{denres}. 

\section{Results}
\label{results}

\subsection{Structure function results}
\label{strfrs}

For the analysis of the structure functions we used the three available {\it
con-noise} maps (see Paper\,I) for the following reasons:

i) The noise has a less predictable effect on the results when using the {\it
con-res} maps. This is caused by the fact that each pixel has a different number
of stars contributing to the $A_V$ value, and hence variable noise. In some
cases only very few stars are used. One drawback of using the {\it con-noise}
maps for this analysis is the variable spatial resolution between pixels. This
is most notable in regions of high extinction. However, we only include pixel
separations above 0.1\,pc (see above), which are generally above the spatial
resolution. Furthermore, we determine the structure functions only for the parts
of the cloud where turbulence dominates, i.e. where $A_V$ is greater than 1\,mag
and less than the cloud's individual star formation threshold (generally below
8\,mag, see Paper\,II). 

ii) The obtained values for $s(2)$ and $s(3)$ are very low when using the {\it
con-res} maps. This could be a result of the generally higher and, more
importantly, variable noise in these maps. With the {\it con-noise} maps we
found much more reasonable values  for $s(2)$ and $s(3)$. We calculate a mean
value for $s(2)$ of 0.8$\pm$0.4 and $s(3)$ of 1.1$\pm$0.6. For individual
clouds, using Taurus  as an example, we find $s(2)$=0.98 and $s(3)$=1.41,
compared to  $s(2)$=0.77 and $s(3)$=1.10 (Padoan et  al.
\cite{2003ApJ...583..308P}). For Perseus we find $s(2)$=1.24 and  $s(3)$=1.75,
compared to $s(2)$=0.83 and $s(3)$=1.18. 

\begin{figure}
\includegraphics[origin=br, angle=270, width=8cm]{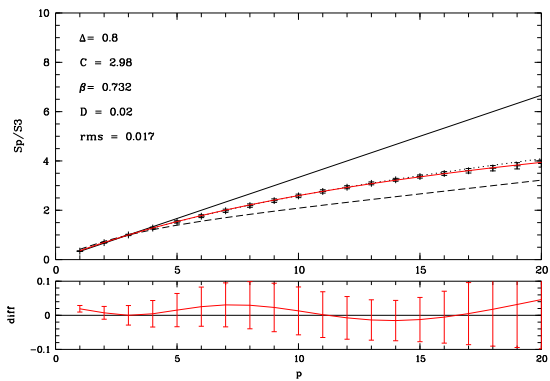} 
\caption{\label{sfeg1} A plot showing the variation of $s(p)/s(3)$ against $p$ 
for the Auriga\,1 cloud and the 49$^{th}$ nearest neighbour map. Our data (+ signs)
is compared against the models of K41 (solid line), SL94 (dotted line) and B02
(dashed line). The best fit values of $\Delta$, $C$, $\beta$, $D$ and the $rms$
value are shown in the upper left corner. A plot of the difference of the fit
from the data against $p$ is shown as a red solid line in the lower panel.}
\end{figure}

Thus, we determine the structure function of every GMC for each of the three
available $A_V$ maps, utilising only extinction values above the self-shielding
limit (1\,mag) and below the star formation threshold of the particular cloud.
Note that every individual extinction value used for these calculations has a
signal-to-noise ratio of at least three (Paper\,I). Hence, the obtained
structure functions are highly reliable, as indicated by the small uncertainties
for $s(p)/s(3)$ (e.g. Fig.\,\ref{sfeg1}).

For each of the three structure functions we determine the best fitting set of
parameters $\Delta$, $C$, $\beta$ and $D$, and  a scatter for each parameter by
considering only the fits with an $rms$ of less than 0.1 (in units of
$s(p)/s(3)$). The final structure function parameters for each cloud are then
determined as weighted averages of the three obtained values. An example plot
showing $s(p)/s(3)$ against $p$ for the Auriga\,1 cloud together with the best
fit is shown in Fig.\,\ref{sfeg1}, where we used the 49$^{th}$ nearest neighbour
map. Similar plots for all clouds are shown in the Appendix\,\ref{indplt}. 

The determined parameters of the structure function ($\Delta$, $C$, $\beta$ and
$D$) for all the clouds and their uncertainties are listed in
Table\,\ref{strrestab}. The $rms$ of the fit is also shown. The values for
$\Delta$ lie in the range 0.45 to 1.08 with a mean of 0.85$\pm$0.16.  Therefore,
even after allowing for uncertainties, the values are more or less in the middle
of those quoted in the literature ($\Delta$=2/3 in e.g. Padoan et al.
\citep{2003ApJ...583..308P}, SL94, B02; $\Delta$=1 in e.g. Schmidt et al.
\citep{2008PhRvL.101s4505S}). 

The co-dimension values $C$ lie between 1.18 and 2.51 with a mean of 
1.92$\pm$0.37. In turn the fractal dimension of our sample of clouds has a value
of 1.08$\pm$0.37, rendering them filament like rather than sheet like.

The values of $\beta$ (the degree of non-intermittency) in our sample range 
from 0.38 to 0.65 with a mean of 0.53$\pm$0.08. The literature usually quotes  a
values for $\beta$ of either 2/3 (SL94) or 1/3 (B02). Therefore, the value for
our GMC sample lies about halfway between these values. 

\begin{table*}
\caption{\label{strrestab} Summary of the results for the 16 investigated GMCs.
The table lists the cloud name (see Paper\,II for coordinate ranges); the best
fitting structure function parameters from Eq.\,\ref{dubeq} and their
uncertainties (see text for details); values of the slopes $s(2)$ and $s(3)$;
average cloud density of material with an $A_V$ above 1\,mag; average cloud
density of material above the star formation threshold (note that
1\,M$_{\odot}/{\rm pc}^3$ corresponds to 40\,cm$^{-3}$); scale of the peak in
the $\Delta$-variance analysis; scale of the peak in the $\Delta$-variance
analysis when applied to the star density maps (a --- denotes that no peak is
visible); the mass spectral index scaling exponent and its uncertainty from the
$\Delta$-variance analysis; weighted correlation coefficient for the
determination of the mass scaling exponent; $\dag$Low density probably caused by
large distance (800\,pc) of cloud and hence partially unresolved higher density
material. $\ddag$High density due to the fact that the cloud is situated in a
region with a general background/foreground extinction. Clouds labeled with
$^1$ are part of the group with steep column density and mass distributions, and
clouds labeled with $^2$ are part of the group with shallow distributions
according to Paper\,II.}
\renewcommand{\tabcolsep}{4pt}
\centering
\begin{tabular}{l|cccccccccc|cc|rr|rr|ccc}
Name & $\Delta$ & $\sigma_{\Delta}$ & $C$ & $\sigma_C$  & $\beta$ &
$\sigma_{\beta}$ & $D$ & $\sigma_D$ & $rms$ & $\sigma_{rms}$ & $s(2)$ & $s(3)$
& $\rho_{av}$ & $\rho_{\rm SF}$ & $\widehat{\sigma_{\Delta^2}^{A_V}}$ & 
$\widehat{\sigma_{\Delta^2}^{\rho}}$ & $\alpha$ & $\sigma_\alpha$ & $r_\alpha$\\
& & & & & & & & & & & & & \multicolumn{2}{c}{[$M_{\odot}/{\rm pc}^3$]} &
[pc] & [pc] & \\ \hline  
$^2$Auriga 1            & 0.77 & 0.07 & 2.05 & 0.55 & 0.59 & 0.13 & 0.95 &  0.55 & 0.05 & 0.01 & 1.15  & 1.66 & 4.9        & 99   & ---   & 10   & 0.36 &  0.07 & 0.980 \\
$^2$Auriga 2            & 0.45 & 0.20 & 1.18 & 0.63 & 0.57 & 0.14 & 1.82 &  0.63 & 0.04 & 0.01 & 0.70  & 0.99 & 12         & 1900 & ---   & ---  & 0.49 &  0.13 & 0.944 \\ 
$^2$Cepheus             & 0.85 & 0.21 & 2.31 & 0.40 & 0.62 & 0.12 & 0.69 &  0.40 & 0.05 & 0.02 & 0.86  & 1.27 & 5.2        & 190  & 16.0  & ---  & 0.18 &  0.10 & 0.984 \\ 
$^1$Chamaeleon 		& 0.96 & 0.06 & 2.51 & 0.34 & 0.62 & 0.04 & 0.49 &  0.34 & 0.05 & 0.02 & 0.72  & 1.05 & 29         & 520  & 3.5   & ---  & 0.63 &  0.09 & 0.949 \\
$^1$Circinus 		& 1.01 & 0.07 & 1.67 & 0.38 & 0.38 & 0.10 & 1.33 &  0.38 & 0.06 & 0.02 & 0.33  & 0.44 & 4.0        & 42   & ---   & ---  & 0.69 &  0.03 & 0.997 \\ 
$^1$Corona Australis 	& 0.88 & 0.07 & 2.28 & 0.45 & 0.60 & 0.08 & 0.72 &  0.45 & 0.04 & 0.02 & 0.66  & 0.93 & 58         & 1500 & ---   & 1.0  & 1.08 &  0.03 & 0.989 \\
$^2$$\lambda$-Ori       & 0.90 & 0.11 & 2.05 & 0.47 & 0.54 & 0.13 & 0.95 &  0.47 & 0.05 & 0.02 & 0.65  & 0.92 & 6.9        & 100  & 12.0  & 4.0  & 0.17 &  0.03 & 0.425 \\ 
$^2$Lupus 1 and 2       & 1.02 & 0.08 & 1.92 & 0.47 & 0.45 & 0.12 & 1.08 &  0.47 & 0.06 & 0.02 & 0.34  & 0.47 & 8.3        & 680  & 1.5   & ---  & 0.23 &  0.10 & 0.644 \\ 
$^2$Lupus 3, 4, 5, 6    & 1.08 & 0.07 & 2.23 & 0.51 & 0.49 & 0.14 & 0.77 &  0.51 & 0.06 & 0.02 & -0.01 & 0.02 & 8.3        & 1500 & 3.8   & ---  & 0.41 &  0.04 & 0.963 \\  
$^2$Monoceros 		& 0.93 & 0.06 & 1.97 & 0.45 & 0.51 & 0.09 & 1.03 &  0.45 & 0.05 & 0.01 & 1.06  & 1.43 & 1.3$\dag$  & 23   & 6.0   & ---  & 0.18 &  0.04 & 0.923 \\ 
$^1$Ophiuchus 		& 0.74 & 0.11 & 1.69 & 0.54 & 0.54 & 0.11 & 1.31 &  0.54 & 0.05 & 0.02 & 1.31  & 1.89 & 18         & 3400 & 1.5   & 1.25 & 0.71 &  0.06 & 0.975 \\ 
$^1$Orion A 		& 0.66 & 0.28 & 1.22 & 0.55 & 0.42 & 0.21 & 1.78 &  0.55 & 0.05 & 0.02 & 1.14  & 1.58 & 12         & 310  & 11.0  & 2.0  & 0.95 &  0.16 & 0.976 \\  
$^1$Orion B 		& 0.83 & 0.09 & 2.05 & 0.58 & 0.57 & 0.11 & 0.95 &  0.58 & 0.05 & 0.02 & 1.33  & 1.91 & 7.0        & 580  & ---   & ---  & 0.52 &  0.05 & 0.950 \\ 
$^1$Perseus 		& 0.79 & 0.12 & 1.75 & 0.51 & 0.53 & 0.08 & 1.25 &  0.51 & 0.04 & 0.01 & 1.24  & 1.75 & 15         & 1600 & ---   & 1.8  & 0.64 &  0.12 & 0.992 \\ 
$^1$Serpens 		& 0.95 & 0.07 & 1.63 & 0.38 & 0.40 & 0.10 & 1.37 &  0.38 & 0.06 & 0.02 & 0.17  & 0.22 & 100$\ddag$ & 860  & ---   & 1.5  & 0.73 &  0.03 & 0.957 \\ 
$^1$Taurus 		& 0.75 & 0.25 & 2.14 & 0.55 & 0.65 & 0.09 & 0.86 &  0.55 & 0.04 & 0.01 & 0.98  & 1.41 & 32         & 6500 & 3.5   & 0.5  & 0.61 &  0.10 & 0.950 \\ 
\end{tabular}
\end{table*}

\subsection{The effect of noise on the structure functions}
\label{noires}

We added noise to a test image and to a selection of the extinction maps of real
clouds to investigate the effect on the structure functions and their fitted
parameters ($\Delta$, $C$, $\beta$, $D$; see Sec.\,\ref{noistr}). The results of
this exercise are shown in Table\,\ref{tesnoires}. 

In the top part of the table we list the values of the structure function
parameters obtained for our test image and the test image plus noise. As one can
see, the artificial cloud corresponds to a structure in agreement with
Kolmogorov type turbulence ($\Delta \approx 0$). This also implies that the
actual values of the other parameters are meaningless (see Eq.\,\ref{dubeq}).
Once our typical noise of 0.28\,mag $A_v$ has been added the parameters change
significantly. The values for $\Delta$ are 0.5 and the structure corresponds to
an almost two-dimensional object ($D$=1.9). When the noise is further increased
no systematic and/or significant changes are observed. This has two
implications: i) Our typical noise can change the structure function parameters
significantly compared to a column density map which is free of noise. ii) Once the image has some observational noise, a small increase in
the noise will not change the values of the determined structure function
parameters significantly or systematically. All changes are well below the
uncertainties of the parameters. This implies that it does not matter which of
our $A_V$ maps we use (and indeed justifies that we average the results from all
three maps), as the variation of a factor of two in the number of stars leads
only to an increase/decrease of the noise in the maps by 33\,\% (caused in part
by the covariance, see Paper\,I).

The bottom part of Table\,\ref{tesnoires} illustrates the latter part of the
above discussion for a real cloud (Corona Australis). With the standard noise in
the 49$^{th}$ nearest neighbour map of 0.28\,mag $A_V$ we obtain the listed
values for the structure function parameters. When increasing the noise by
another small amount (at maximum 25\,\%), again we do not find any significant
and/or systematic changes in the determined structure function parameters.

\begin{table}
\caption{\label{tesnoires} Structure function parameters obtained by our test
with increased noise. We list the noise added to the image (in units of
0.28\,mag $A_V$, the 1\,$\sigma$ noise in our 49$^{th}$ nearest neighbour map)
and the structure function parameters as well as the $rms$. {\bf Top:} Results
when adding noise to our artificial test image. {\bf Bottom:} Example for the
test with a real cloud, in this case Corona Australis. We only used the
49$^{th}$ nearest neighbour map, and the original image is thus represented by
the first line. The noise values for this case hence represent the total noise,
not just the added noise.}
\renewcommand{\tabcolsep}{3pt}
\centering
\begin{tabular}{c|cccccccccc}
Noise & $\Delta$ & $\sigma_{\Delta}$ & $C$ & $\sigma_C$ & $\beta$ & 
$\sigma_{\beta}$ & $D$ & $\sigma_D$ & $rms$ & $\sigma_{rms}$ \\ 
$\left[\sigma\right]$ & & & & & & & & & & \\ \hline
\multicolumn{11}{l}{artificial test cloud} \\
0.0000 & 0.02 & 0.01 & 2.03 & 0.58 & 0.99 & 0.01 & 0.97 & 0.58 & 0.01 & 0.01 \\ 
1.0000 & 0.50 & 0.06 & 1.18 & 0.41 & 0.55 & 0.11 & 1.82 & 0.41 & 0.04 & 0.01 \\ 
1.0625 & 0.50 & 0.06 & 1.21 & 0.42 & 0.55 & 0.10 & 1.79 & 0.42 & 0.04 & 0.01 \\ 
1.1250 & 0.48 & 0.06 & 1.09 & 0.40 & 0.53 & 0.12 & 1.91 & 0.40 & 0.05 & 0.01 \\ 
1.1875 & 0.48 & 0.06 & 1.17 & 0.43 & 0.56 & 0.11 & 1.83 & 0.43 & 0.04 & 0.01 \\ 
1.2500 & 0.44 & 0.06 & 0.96 & 0.38 & 0.49 & 0.14 & 2.04 & 0.38 & 0.05 & 0.01 \\ \hline
\multicolumn{11}{l}{Corona Australis} \\
1.0000 & 0.85 & 0.03 & 2.64 & 0.25 & 0.68 & 0.02 & 0.36 & 0.25 & 0.09 & 0.01 \\ 
1.0625 & 0.83 & 0.06 & 2.35 & 0.41 & 0.64 & 0.04 & 0.65 & 0.41 & 0.03 & 0.01 \\ 
1.1250 & 0.83 & 0.06 & 2.39 & 0.41 & 0.65 & 0.04 & 0.61 & 0.41 & 0.03 & 0.01 \\ 
1.1875 & 0.86 & 0.06 & 2.42 & 0.40 & 0.64 & 0.04 & 0.58 & 0.40 & 0.03 & 0.01 \\ 
1.2500 & 0.92 & 0.05 & 2.57 & 0.33 & 0.64 & 0.03 & 0.43 & 0.33 & 0.04 & 0.01 \\ 
\end{tabular}
\end{table}

\subsection{$\Delta$-variance results}
\label{delvrs}

\begin{figure}
\includegraphics[width=8cm]{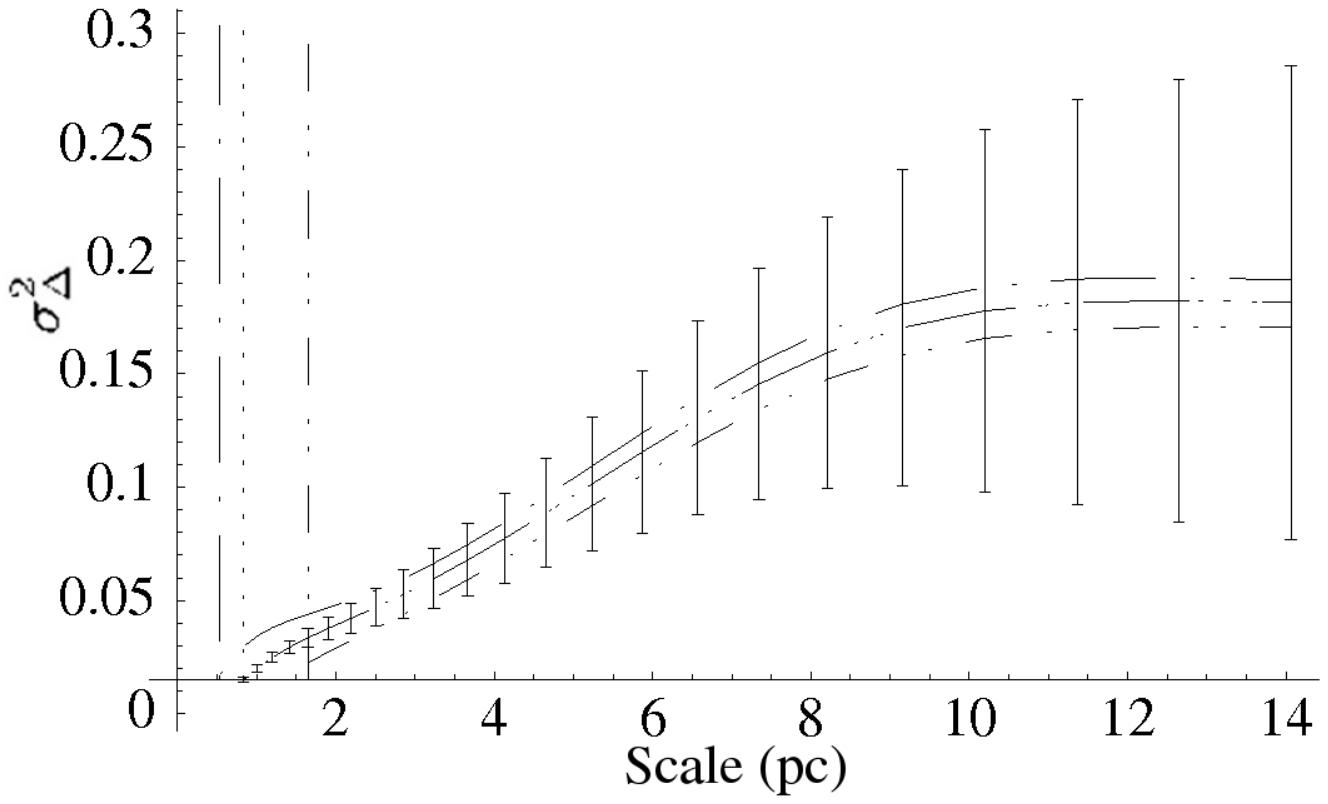} 

\caption{\label{dvex1} Example of the delta variance $\sigma^2_\Delta$ for 
Orion\,A. We show the results for all three available $A_V$ maps ({\it dash-dot}
25$^{th}$ nearest neighbours, {\it dotted} 49$^{th}$ nearest neighbour, {\it
dash-dot-dot} 100$^{th}$ nearest neighbours). The corresponding vertical lines
indicate the spatial resolution limit for each map, below which the values are
meaningless. The typical scale of the structure in this cloud can be identified
by the peak at about 11\,pc. For clarity, error bars are only overplotted on the
49$^{th}$ nearest neighbour data. The weighted correlation coefficient for the
determined slope is 0.976.}

\end{figure}

We applied the $\Delta$-variance technique (as described in  Sec.\,\ref{delmet})
to all GMCs of our sample. As an example we show the results of the
$\sigma^2_\Delta$ plot for Orion\,A in Fig.\,\ref{dvex1}. Similar plots for  all
the individual clouds can be found in the Appendix\,\ref{indplt}. These plots
also contain a cut-out of the 49$^{th}$ nearest neighbour $A_V$ map, as well as
the structure function of the cloud, determined from this map.

The statistical error bars for the $\Delta$-variance in some clouds appear
rather large. This has also been recognised by Ossenkopf et al.
\cite{2008A&A...485..719O}. However, they also note that (quote): "In spite of
the large error bars, the general scaling behaviour can be accurately traced."
Since we will measure the mass spectral index scaling exponents $\alpha$ in
these diagrams we determined the weighted correlation coefficients $r_\alpha$
(listed in the last column of Table\,\ref{strrestab}). They show, except for two
clouds ($\lambda$-Ori, Lupus\,1, 2) that there is a very good correlation, with
$r_\alpha$ generally significantly larger than 0.9.

Using our {\it con-noise} maps we investigated the positions of the peaks in the
$\Delta$-variance plots. These peaks reveal the scale at which a change in the
structure occurs, e.g. the  size of the cloud, the length of filaments etc..
These results are shown in Table\,\ref{strrestab}. We find peaks in some of the 
clouds (Cepheus, Chamaeleon, $\lambda$-Ori, Orion\,A, Lupus, Monoceros, Taurus)
while for the remaining clouds no clear-cut peak can be identified. The
identified peaks appear over a range of scales from  1.5\,pc (Lupus\,1 and 2 and
Ophiuchus) to 16\,pc (Cepheus). 

Using the star density instead of the column density maps, we find peaks in the
$\Delta$-variance for the clouds Auriga\,1, Circinus, Corona Australis,
$\lambda$-Ori, Ophiuchus, Orion\,A, Perseus, Serpens and Taurus. In general, the
peaks occur at smaller scales (from  0.5\,pc in Taurus to 10\,pc in
Auriga\,1). However, only for four clouds ($\lambda$-Ori, Ophiuchus, Orion\,A,
Taurus) do we detect a clear peak in both maps. In each case we find that the
peak determined when using the column density map is larger, indicating that
the star density maps trace smaller scales when the $\Delta$-variance technique
is applied to them.

By means of the {\it con-noise} maps  we also measure the mass spectral index
scaling exponents $\alpha$ for each cloud. These values and uncertainties are
listed in the last two columns of Table\,\ref{strrestab}. The listed values are
the mean of the results for the three available column density maps. The values
for $\alpha$ range from 0.17 ($\lambda$-Ori) to 1.08 (Corona Australis) with a
mean for all clouds of 0.54$\pm$0.27.

\subsection{The effect of noise on the $\Delta$-variance}
\label{noiresdvar}

Similar to the investigations of the structure function, we determined the
influence of noise on our analysis of the cloud structure with the
$\Delta$-variance method. The same noise increments as for the structure
function analysis were added to the extinction maps and the change to the peak
position of the $\Delta$-variance and the mass spectral index scaling exponent
were determined.

Neither the peak position nor the mass spectral index seem to be significantly
and/or systematically influenced, if an additional noise of the level discussed
in Sect.\,\ref{noistr} is added to the extinction maps. This applies to both,
the use of our artificially generated image as well as the real cloud. Hence the
obtained values can be considered robust.

\subsection{Molecular cloud density}
\label{denres}

The determined average densities of the entire GMCs, as well as the densities of
the material above the star formation threshold can be found in
Table\,\ref{strrestab}. 

The average densities for the entire cloud range from 1.3\,$M_{\odot}/pc^3$ (for
Monoceros) to 100\,$M_{\odot}/pc^3$ (for Serpens). The low value for Monoceros
might be due to small scale structures not being detected because of the
distance to the cloud. For the Serpens cloud the apparently high density is due
to a general offset in the extinction values (either foreground or background to
the cloud). When not considering the very low and very high values, we find a
mean value for our sample of about 15\,$M_{\odot}/pc^3$, which corresponds to
600\,cm$^{-3}$. 

The densities of material above the star formation threshold are naturally
higher. We find a range from 23\,$M_{\odot}/pc^3$ (Monoceros) to
6500\,$M_{\odot}/pc^3$ (Taurus). The median value is about 750\,$M_{\odot}/pc^3$
or 3\,$\cdot$\,10$^4$\,cm$^{-3}$. Here the values for the individual clouds
should be considered with great care, as we find larger values for clouds with
smaller distances, indicating the non-detection of higher column density
material in more distant clouds (due to limited spatial resolution) and thus an
underestimate of the density. 

\section{Discussion}
\label{discuss}

\subsection{Star Forming properties}

To investigate possible influences of cloud structure on the star formation
properties we evaluate the current and potential future amount of star formation
of our sample. 

i) In Paper\,II (e.g. bottom left panel of Fig.\,4) we found that our sample of
clouds can be divided into two groups when considering their column density and
mass distribution in the low extinction regions (A$_V$ above 1\,mag and below
the star formation threshold). We indicate which cloud belongs to which group in
Table\,\ref{strrestab}. Clouds in group\,1 have a shallow column density and
mass distribution while clouds in group\,2 have a steeper distribution. Hence,
group\,1 members have a larger amount of material at high column densities, and
thus more mass potentially available for star formation. One finds that the
average fraction of mass potentially involved in star formation is about two to
three times higher for clouds in group\,1 (see Table\,4 in Paper\,II).

ii) We use literature data to estimate the current number of young stars and
mode of star formation for each cloud. In particular the work by Kainulainen et
al. \cite{2009A&A...508L..35K} and Reipurth
\cite{2008hsf1.book.....R,2008hsf2.book.....R} have been used. It turns out that
clouds in group\,1 contain typically more than 100 young stellar objects and the
majority of these clouds form at least one cluster of stars. Clouds in group\,2
on the other hand, typically only form a few tens of stars and no clusters. 

Thus, group\,1 contains clouds which intensely form stars, preferably in a
clustered mode, while clouds in group\,2 show weaker star formation activity
mostly in a distributed mode. Taurus, without a known cluster of young stars,
seems to be an exception as it belongs to group\,1.

\subsection{Structure functions}
\label{strfan}

For our structure function analysis we selected only pixel values between
the self shielding column density threshold of 1\,mag $A_V$ and the individual
star formation threshold for each cloud (see Paper\,II), as this range
represents the part of the clouds where turbulence is expected to dominate the
structure. We  calculated the structure functions for each cloud using the three
{\it con-noise} maps and a spatial scale range from 0.1\,pc to 1.0\,pc
(traceable for all clouds). 

Using the data shown in Table\,\ref{strrestab} we investigated possible
correlations between the parameters $\Delta$, $C$ against the molecular cloud 
densities ($\rho_{av}$ and $\rho_{SF}$) as well as a range of other parameters 
from Table\,4 of Paper\,II. No correlations could be found. The analysis was 
also performed by considering two separate groups of clouds. There are no
significant differences for any of the structure function parameters between
the two groups. Hence, our determined structure functions are not able to
distinguish between clouds that form stars in clusters or in a distributed mode.

One might expect that the values of the co-dimension $C$ increase with distance
(i.e. the fractal dimension $D$ is decreasing), due to the change in appearance
of the cloud. More distant clouds will appear to have a more simplistic
structure due to details not being resolved. However, in our analysis no such
trend could be found. This can be attributed to the fact that we determine the
structure functions in the scale range 0.1\,pc to 1.0\,pc, which is resolved for
all investigated clouds. 

Our mean value for all clouds of $\Delta$\,=\,0.85$\pm$0.16 is much higher than
the values of $\Delta$\,=\,0 expected by K41. Furthermore, it is inbetween the
predictions of $\Delta$\,=\,2/3 (SL94, B02) and $\Delta$\,=\,1.0 (S08). However,
the scatter in our results means the value could be as low as $\Delta$\,=\,0.69
or as high as $\Delta$\,=\,1.01. For the parameter $C$ our average of
1.92$\pm$0.37 is close to the SL94 value of $C$\,=\,2.0, while B02 theorised a
value of $C$\,=\,1.0. In S08 $C$ is found to be $\approx$\,1.1 for compressive
forcing and $\approx$\,1.5 for solenoidal forcing, both below our average value
for the co-dimension. Our average is closer to the predicted value for
solenoidal forcing (in particular considering the uncertainties). Hence, we
conclude that the average cloud of our sample has a structure function hinting
to solenoidal forcing of the turbulent field, rather than compressive driving.

 %  
 %  Judging, however, the individual values for the structure function parameters,
 %  our sample of clouds could be governed by a variety of different turbulent
 %  fields. None of the clouds, however, fits perfectly to the models of K41, SL94,
 %  B02 or S08. Rather we see a mix of turbulent driving mechanisms, the precise
 %  nature depending on the local environment of the cloud.
 %  

 %  
 %  However, this interpretation needs to be considered with considerable caution. 
 %  It is not clear at all if the differences in the observed values can be
 %  attributed to different turbulent fields in the clouds, or simply represent
 %  effects of the viewing angle and/or other biases. For example the random
 %  distribution of background stars to the cloud will influence the determination
 %  of the $A_V$ map from the real column density of the cloud. Furthermore, the
 %  conversion of real cloud column density into an extinction map (in which the
 %  structure function is determined) might change the structure function
 %  parameters. Hence a comparison of the determined parameters to model calculation
 %  to infer the type of turbulence in the cloud needs to be done in a more
 %  realistic way, i.e. one should consider the whole process of the extinction map
 %  determination to ensure that no changes of parameters occur. We will detail this
 %  in the forthcoming paper Froebrich et al. (in prep.).
 %  

\subsection{$\Delta$-variance}
\label{delvan}

Using the $\Delta$-variance technique, we identified peaks in plots of
$\sigma^2_\Delta$ against $\Delta r$ to quantify the scale of structure in our
investigated GMCs. We were only able to find peaks in $\sigma^2_\Delta$ for
about half the clouds. In the remainder no peaks could be identified. This could
be caused by the fact that there are no dominant scales in those clouds, or that
the size of the map around each cloud was too small (due to neighbouring
clouds). Note that the $\Delta$-variance will only detect dominant scales if
they are significantly smaller than the map size. 

Scales identified average at about 5\,pc. These generally correspond to the
width of a filament or sheets in the cloud rather than the length or size of the
entire cloud. In the region Lupus\,3, 4, 5, 6 the peaks correspond to the length
of the different filaments.  When we perform the $\Delta$-variance analysis with
the star density maps, dominant scales are found as well for about half the
clouds. These scales are smaller and average at about 2\,pc. This indicates that
the star density maps trace more compact, higher extinction regions, generally
the width of dense filaments.

For the mass spectral index scaling exponent $\alpha$, Federrath et al.
\cite{2010A&A...512A..81F} found values of $\alpha$\,=\,0.55 for solenoidal
forcing of turbulence and $\alpha$\,=\,1.34 for compressive driving. We find a
considerable variation in the measured values for $\alpha$ amongst our clouds.
All values are below the predicted amount for pure compressive driving. However,
the mean for all clouds is $\alpha$\,=\,0.54$\pm$0.27, hence very close to the
predictions for pure solenoidal driving of turbulence. 

When considering the average values of $\alpha$ for our two groups of clouds,
significant differences can be found. While the cluster forming clouds have an
average of $\alpha$\,=\,0.73$\pm$0.18, the clouds with a more distributed star
formation mode have a mean of $\alpha$\,=\,0.29$\pm$0.13 (or, when excluding the
clouds with low correlation coefficient, $\alpha$\,=\,0.32$\pm$0.14). Based on
the predictions from Federrath et al. \cite{2010A&A...512A..81F}, the larger
value of $\alpha$ for the cluster forming clouds indicates an enhanced component
of compressive driving compared to the clouds with a more distributed mode of
star formation. We refrain from trying to estimate a more quantitative statement
about the amount of contribution from compressive driving in the cluster forming
clouds, since the uncertainties and scatter in our data are clearly large.

\section{Conclusions}
\label{conclusions}

We present the homogeneous investigation of a selection of 16 nearby Giant
Molecular Clouds. Utilising near infrared extinction maps made from 2MASS data
with the near infrared colour excess technique, we determine structure functions
and perform a $\Delta$-variance analysis. The same clouds as in Froebrich \&
Rowles \cite{2010MNRAS.406.1350F} were investigated. They are all reasonably
nearby, away from the Galactic Plane, and to the best of our knowledge there is
only one cloud along each line of sight. 

Using our extinction maps we identify two groups of clouds based on the slope
of their column density and mass distributions of the turbulence dominated (low
A$_V$) part of the cloud. We find that clouds with shallower mass distributions
form stars preferably in a clustered mode, while clouds with steep mass
distributions form fewer stars and show a more distributed mode of star
formation.

Structure functions determined for all clouds homogeneously within a range from
0.1\,pc to 1.0\,pc cannot be used to distinguish clouds with different star
formation modes. Comparing the structure function parameters to model
calculations suggests that the turbulence in the investigated clouds is governed
preferably by solenoidal forcing.

Our results of the $\Delta$-variance analysis also indicate that the average
cloud in our sample is governed by solenoidal forcing. However, clouds which
form a large number of stars in clusters have an enhanced component of
compressive driving of the turbulent field, in comparison to clouds with
isolated star formation. Hence, compressive driving seems to lead to a more
clustered mode of star formation.

To quantify these qualitative findings, i.e. to determine the fraction of
solenoidal and compressive driving in each cloud, a more detailed comparison of
the numerical simulations with the observational data needs to be performed. In
particular viewing angle and resolution/distance effects in the model data need
to be investigated.

\section*{acknowledgements}
\label{acks}

JR acknowledges a University of Kent scholarship. This publication makes use 
of data products from the Two Micron All Sky  Survey, which is a joint project 
of the University of Massachusetts and the Infrared Processing and Analysis 
Center/California Institute of Technology, funded by the National Aeronautics 
and Space Administration and the National Science Foundation.

\clearpage
\newpage
\begin{appendix}
\section{Structure function results}
\subsection{Individual clouds}
\label{indplt}

\begin{figure*}
\centering
\beginpicture
\setcoordinatesystem units <-3.8mm,3.8mm> point at  0 0
\setplotarea x from 173 to 156 , y from -12 to -3
\put {\includegraphics[width=6.5cm, height=5cm]{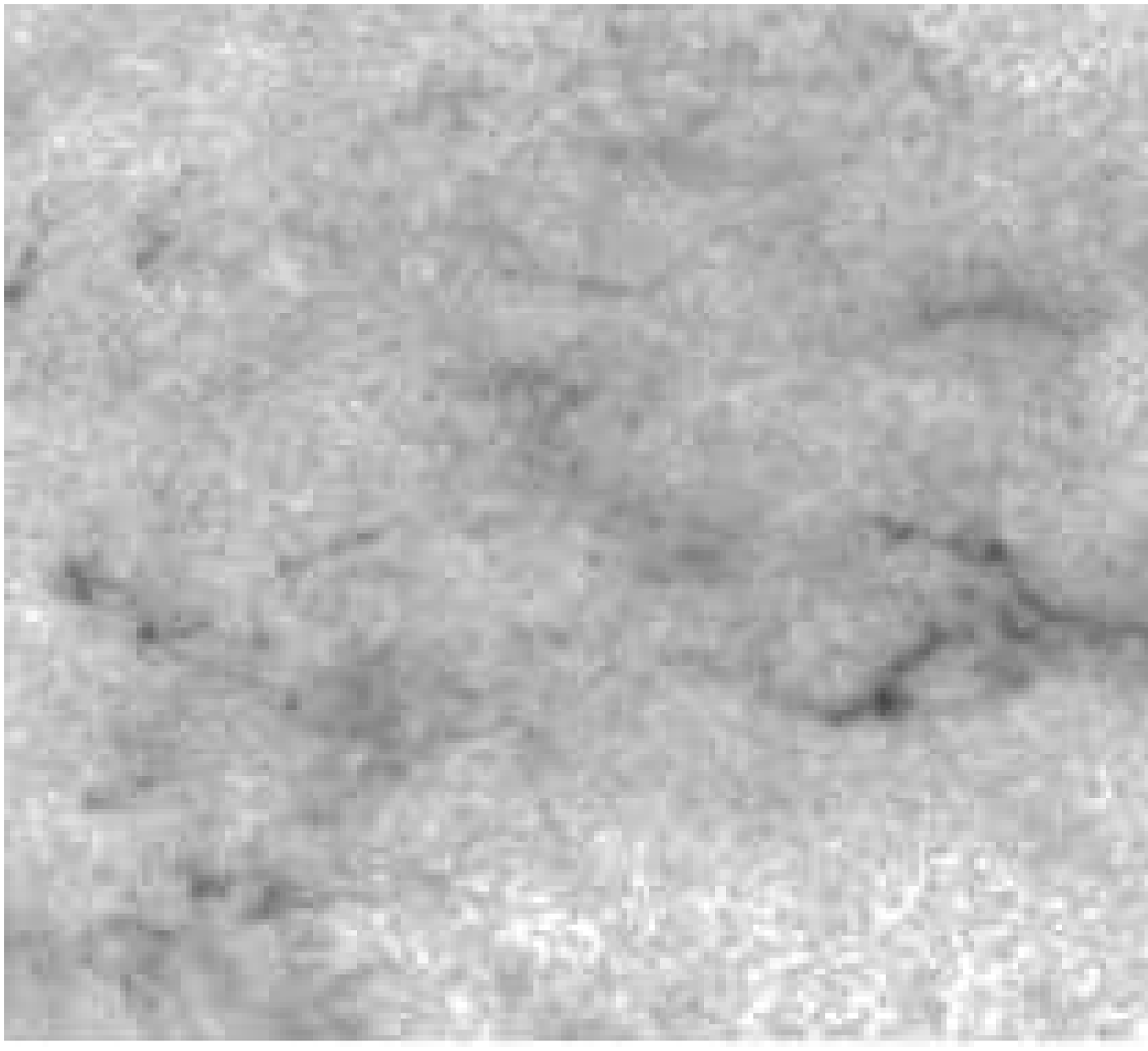}} at 164.5 -5.5
\axis left label {\begin{sideways}b [\degr]\end{sideways}}
ticks in long numbered from -12 to -3 by 5
      short unlabeled from -12 to -3 by 1 /
\axis right label {}
ticks in long unlabeled from -12 to -3 by 5
      short unlabeled from -12 to -3 by 1 /
\axis bottom label {l [\degr]}
ticks in long numbered from 156 to 173 by 5
      short unlabeled from 156 to 173 by 1 /
\axis top label {}
ticks in long unlabeled from 156 to 173 by 5
      short unlabeled from 156 to 173 by 1 /
\put {\line(1,0){27.7}} at 171 -10
\put {20 pc} at 171 -11
\put {\includegraphics[width=8.0cm]{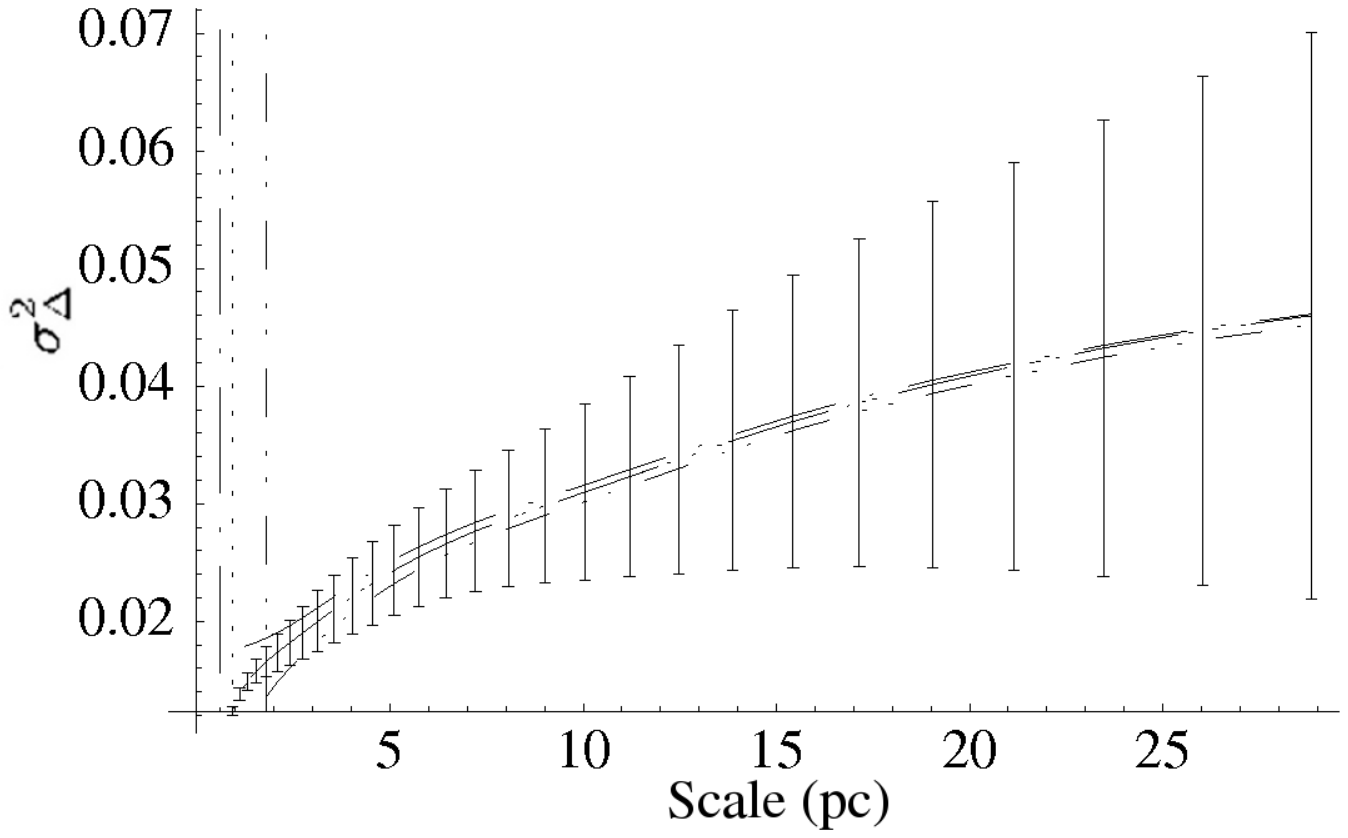}} at 140. -8.
\endpicture
\includegraphics[origin=br, angle=270, width=8.0cm]
{images/jr_result_plot_0001.pdf} \hfill
\includegraphics[width=8.0cm]{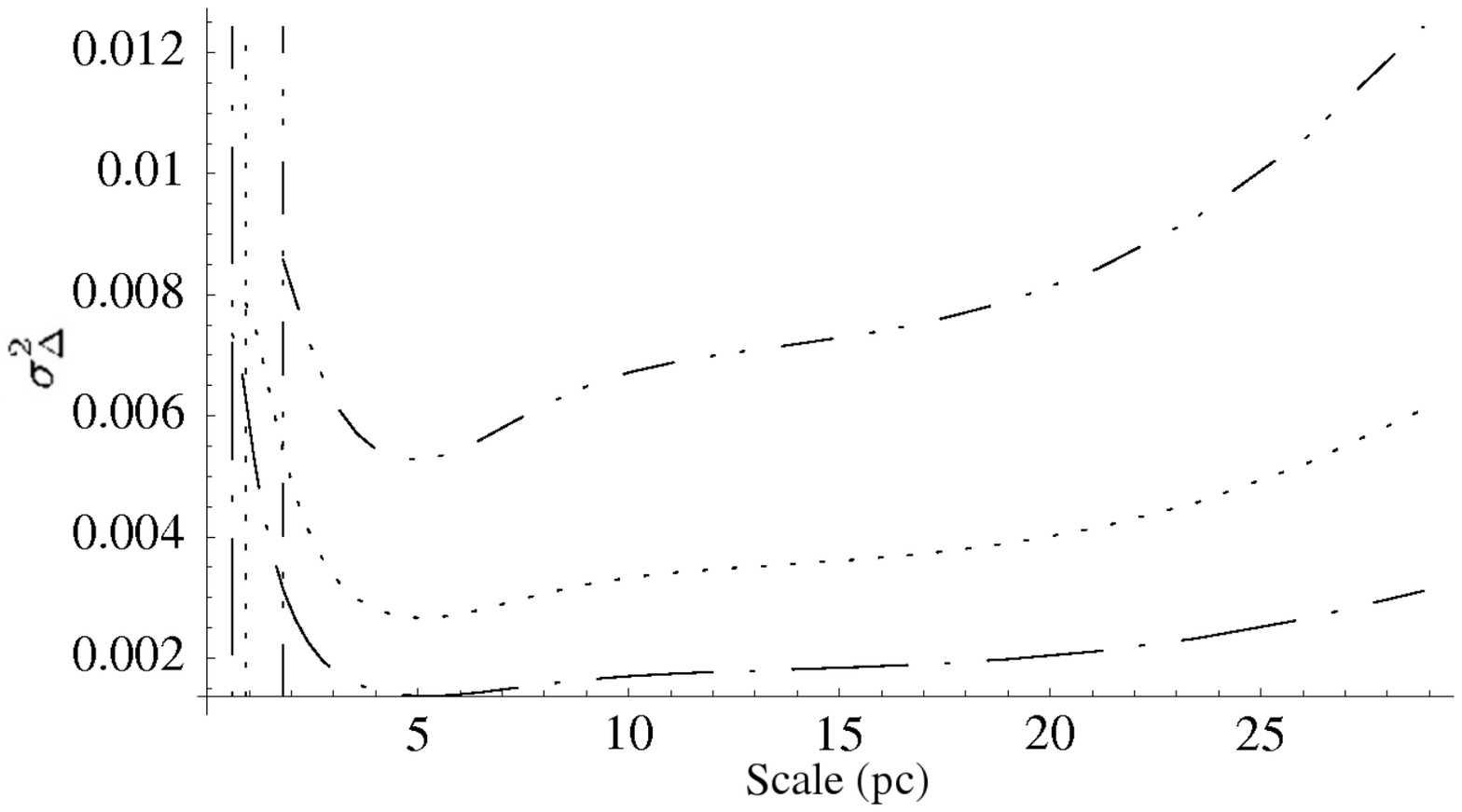} \\

\caption{\label{fig_aur1} Sample of the Appendix. All other figures are
available online only . {\bf Top left:} An extraction of the $A_V$ map (using
the nearest 49 stars) around Auriga\,1. Gray scales are square root scaled from
0\,mag (white) to 15\,mag optical extinction (black). The size of the image at
the distance of the cloud is 133.5\,pc by 70.7\,pc. {\bf Bottom left:} A plot
showing the structure function $s(p)/s(3)$ against $p$. Our data (crosses) is
compared against the models of K41 (black solid line), SL94 (black dotted line)
and B02 (black dashed line). The best fit is shown as red solid line and its
parameter values ($\Delta$, $C$, $\beta$, $D$, and the $rms$) are listed in the
upper left of the panel. The difference of the fit to the data against $p$ is
also shown as red solid line in the bottom of the panel. {\bf Top right:}
$\Delta$-variance calculated using our {\it con-noise} maps. The {\it dash-dot}
line denotes the $A_V$ map with the nearest 25 stars used, the {\it dotted} line
denotes the $A_V$ map with the nearest 49 stars used, and the {\it dash-dot-dot}
line denotes the $A_V$ map with nearest 100 stars used. For clarity, error bars
are only shon for the 49 stars data. {\bf Bottom right:} $\Delta$-variance
calculated using our star density maps. The {\it dash-dot}, {\it dotted}, and
{\it dash-dot-dot} lines are as in the upper right panel.}

\end{figure*}

\end{appendix}

\end{document}